\documentclass[a4paper,aip,apl,amsmath,amssymb,reprint,superscriptaddress]{revtex4-1}
\pdfoutput=1
\usepackage{subcaption}
\usepackage{graphicx}
\usepackage{amsmath}
\usepackage{dcolumn}
\usepackage{bm}
\usepackage{bbold}
\usepackage{color}
\usepackage{amsfonts}
\usepackage{amssymb}
\usepackage{mathrsfs}
\usepackage{tabularx}
\usepackage{braket}
\usepackage{mathtools}
\usepackage{soul}			

\newcommand{\deriv}[2][t]{\frac{\mathrm{d}#2}{\mathrm{d}#1}}
\newcommand{\derivtwo}[2][t]{\frac{\mathrm{d^{2}}#2}{\mathrm{d}#1^{2}}}

\begin{document} 
\title{Experimental Implementations of Cavity-Magnon Systems: from Ultra Strong Coupling to Applications in Precision Measurement}

\author{Graeme Flower}
\email{21302579@student.uwa.edu.au}
\affiliation{ARC Centre of Excellence for Engineered Quantum Systems, Department of Physics, University of Western Australia, 35 Stirling Highway, Crawley WA 6009, Australia}

\author{Maxim Goryachev}
\affiliation{ARC Centre of Excellence for Engineered Quantum Systems, Department of Physics, University of Western Australia, 35 Stirling Highway, Crawley WA 6009, Australia}

\author{Jeremy Bourhill}
\affiliation{ARC Centre of Excellence for Engineered Quantum Systems, Department of Physics, University of Western Australia, 35 Stirling Highway, Crawley WA 6009, Australia}

\author{Michael E. Tobar}
\email{michael.tobar@uwa.edu.au}
\affiliation{ARC Centre of Excellence for Engineered Quantum Systems, Department of Physics, University of Western Australia, 35 Stirling Highway, Crawley WA 6009, Australia}

\date{\today}

\begin{abstract}
Several experimental implementations of cavity-magnon systems are presented. First an Yttrium Iron Garnet (YIG) block is placed inside a re-entrant cavity where the resulting hybrid mode is measured to be in the ultra strong coupling regime. When fully hybridised the ratio between the coupling rate and uncoupled mode frequencies is determined to be $g/\omega=0.46$. Next a thin YIG cylinder is placed inside a loop gap cavity. The bright mode of this cavity couples to the YIG sample and is similarly measured to be in the ultra strong coupling regime with ratio of coupling rate to uncoupled mode frequencies as $g/\omega=0.34$. A larger spin density medium such as lithium ferrite (LiFe) is expected to improve couplings by a factor of 1.46 in both systems as coupling strength is shown to be proportional to the square root of spin density and magnetic moment. Such strongly coupled systems are potentially useful for cavity QED, hybrid quantum systems and precision dark matter detection experiments. The YIG disc in the loop gap cavity, is, in particular, shown to be a strong candidate for dark matter detection. Finally, a LiFe sphere inside a two post re-entrant cavity is considered. In past work it was shown that the magnon mode in the sample has a turnover point in frequency\cite{CMP_Life}. Additionally, it was predicted that if the system was engineered such that it fully hybridised at this turnover point the cavity-magnon polariton (CMP) transition frequency would become insensitive to both first and second order magnetic bias field fluctuations, a result useful for precision frequency applications. This work implements such a system by engineering the cavity mode frequency to near this turnover point, with suppression in sensitivity to second order bias magnetic field fluctuations shown.
\end{abstract} 

\maketitle
\section*{Introduction}
Magnonic systems have been of considerable interest recently. Applications of such systems range from quantum information processing \cite{magnonDataProcessing, magnonSpintronics, NVspin} and coherent conversion of microwave to optical frequency light \cite{microlightconv1,microlightconv2}, to microwave components in the form of filters, circulators, isolators and oscillators. Additionally, such systems are used in the study of hybrid quantum systems\cite{magnonMeetsQubit, resolvingmagnons}, Quantum electrodynamics (QED) \cite{CQED1,CQED2,CQED3}, and direct detection of dark matter \cite{me, quaxnew, quaxprop, darkmatteratom, casper}. In the context of dark matter detection, it has been shown that strongly coupled cavity-magnon systems are useful for expanding the range of detectable dark matter masses\cite{me}. Typically, the material of choice for these experiments is Yttrium Iron Garnet (YIG) due to its low magnonic and photonic loss, and high spin density, however, other ferrimagnetic materials are often considered for study such as lithium ferrite (LiFe) \cite{CMP_Life} and ${\mathrm{Cu}}_{2}{\mathrm{OSeO}}_{3}$ \cite{copperferrite}. LiFe has been of recent interest for use in hybrid cavity-magnon systems due to is higher spin density when compared to YIG. It was additionally shown to have a turnover point in its frequency as a function of DC magnetic field \cite{CMP_Life}. This is of interest as the cavity-magnon polariton (CMP) transition frequency (difference frequency of hybrid modes) was predicted to become insensitive to both first and second order magnetic field fluctuations if the system were to be fully hybridised at this turnover point. This is useful for applications which require precision frequency measurements, as this would reduce the effect of fluctuations of magnetic field biasing and is similar to a double magic point atomic clock transition\cite{magicClock}. One aim of this work was to implement this system to demonstrate suppression of second order magnetic field fluctuations in the CMP transition frequency.\\

Often, in the study of hybrid quantum systems, a perturbative approach, such as the rotating wave approximation (RWA), is used to analyse dynamics. This is valid when the ratio of coupling rate to bare mode frequency is small ($g/\omega\ll 1$) and predicts that as the coupling rate increases between subsystems, there is improved coherence of information exchange and a larger spontaneous emission rate \cite{lightmatterDSC}. A goal in the study of hybrid quantum systems is thus to achieve stronger couplings between the component systems. At larger couplings, however, the RWA begins to break down and new dynamics appear. The ultra-strong coupling (USC) regime, occurs when $g/\omega> 0.1$, and, in the context of light matter couplings has been shown to lead to interesting new observations including the dynamical Casimir effect \cite{casmir1, casmir2}, super-radiant phase transitions \cite{superRad1, superRad2, superRad3}, and ultra-efficient light emission \cite{efflight1, USCtheory2, efflight2}. Further interesting dynamics are expected to appear in the deep-strong coupling (DSC) regime, where $g/\omega> 1$, such as the counter intuitive result that energy exchange between component subsystems saturates and then drops off when moving from USC to DSC. This leads to a saturation followed by reduction in the spontaneous emission rate for larger couplings \cite{lightmatterDSC}. Where the perturbative approach to solving dynamical equations breaks down in the USC and DSC regimes, new theoretical approaches have appeared \cite{USCtheory1,USCtheory2,DSCqubittheory}. The USC regime has been explored experimentally in various applications from coupled photons and superconducting qubits\cite{USCqubit1, USCqubit2}, to cavity-magnonic systems \cite{CQED1, USCmagnons, strongmagnons, CQED3}, to other forms of light matter coupling \cite{USClightmatter1, USClightmatter2, USClightmatter3, USClightmatter4, USClightmatter5}. The DSC regime has also now been demonstrated experimentally in superconducting circuits\cite{QbDscExp1, QbDscExp2}, and terahertz light-Landau polariton couplings in nanostructure metamaterials\cite{LmDscExp}. A comprehensive review of such phenomenon was recently published \cite{USCreview}.\\

This work considers some experimental implementations of the USC regime in cavity-magnon polaritons through cavity engineering. The measurement set-up in all cases is to place the cavity, made from oxygen free copper, inside a solenoidal superconducting magnetic and cooled inside a Dilution refrigerator (see below for specific temperatures). The frequency response in transmission of the system is then measured as a function of DC magnetic field. This procedure is explained in more detail in previous works\cite{pastWork1, pastWork2}.

\section{Coupling in Cavity-Magnon systems}
In the absence of a general model for magnonic systems of arbitrary geometries it is useful to consider the case of a cavity mode coupled to a uniform precession magnon mode. The Hamiltonian of the cavity-magnon system consists of its cavity and magnon parts, $H_c$ and $H_m$ respectively, as well as magnon-cavity interaction, $H_\textrm{int}$:
\begin{multline}
\begin{aligned}
H = &H_c + H_m + H_\textrm{int}\\
H = &\hbar\omega_c c^\dagger c + \hbar\omega_m b^\dagger b + \hbar g_{cm}^{x}(c+c^\dagger)(b+b^\dagger) \\
&+ \hbar g_{cm}^{y}(c+c^\dagger)(b-b^\dagger),
\end{aligned}
\end{multline}
where $c^\dagger$ ($c$) is a creation (annihilation) operator for photon, $b^\dagger$ ($b$) is a magnon creation (annihilation) operator, $\omega_c$ is the cavity frequency, $\omega_m$ is magnon frequency, $g_{cm,x}$ ($g_{cm,y}$) is the cavity magnon coupling rate associated with overlap of x (y) directional field and $\hbar$ is the reduced Planck's constant. It is assumed that the DC bias magnetic field is in the z direction and the particular choice of x and y directions is shown later to be arbitrary. These expressions can been found from first principles \cite{cavitymagnon, magnonkerr} where the interaction term is derived in the supplementary materials. \\

Eigen-frequencies, $\omega_\pm$, of this coupled mode system are (neglecting dissipation):
\begin{equation}
\omega_\pm = \sqrt{\frac{\omega_c^2+\omega_m^2}{2} \pm \sqrt{\Big(\frac{\omega_c^2-\omega_m^2}{2}\Big)^2+4\omega_c\omega_mg_{cm}^2}},
\label{2mode}
\end{equation}
where the coupling rate $g_{cm}^2=(g_{cm}^x)^2-(g_{cm}^y)^2$, can be written as (see supplementary materials):
\begin{align}
g_{cm} &= \frac{\gamma}{2}\eta\sqrt{\frac{\mu_0S\hbar \omega_c}{V_m}}, \label{uniform}\\
\eta &= \sqrt{\frac{\big(\int_{V_m}\mathbf{H}\cdot\hat{\mathbf{x}}\mathrm{d}V\big)^2 + \big(\int_{V_m}\mathbf{H}\cdot\hat{\mathbf{y}}\mathrm{d}V\big)^2}{V_m\int_{V_c}|\mathbf{H}|^2\mathrm{d}V}},
\end{align}
where $\eta$ is a form factor ranging from $0$ to $1$, and $S$ is the total spin number of the macrospin operator. Note, $S$ is determined by $S=\frac{\mu}{g\mu_B}N_s$, where $\mu$ is the magnetic moment of the magnetic sample, $\mu_B$ is the Bohr magneton, $g$ is the g-factor ($g=2$) and $N_s$ is the number of spins in the sample (given by $N_s=n_sV_m$ with $n_s$ as spin density and $V_m$ as volume). When $\eta=0$ none of the cavity $\mathbf{H}$-field is perpendicular to the external DC field and contained in the magnetic sample. In contrast, when $\eta=1$ all of the cavity $\mathbf{H}$-field is perpendicular and in the sample. It can be noted from the form of this expression that the coupling rate isn't directly dependent on the volume of the magnetic sample (as $S\propto V_m$), contrary to some past papers \cite{CQED3, CQED2}, where they have claimed $g_{cm}\propto\sqrt{N_s}$, implying $g_{cm}\propto\sqrt{V_m}$. Instead, increasing the volume of the magnetic sample in the cavity leads to larger form factor, $\eta$, as more of the cavity field is contained in the magnetic material. An increase in this filling factor can similarly be achieved, however, by cavity design without increasing the volume of the magnetic sample. For example, the use of two post re-entrant cavities has been shown to produce large coupling rates for small magnetic volumes\cite{CQED1}. In this work, it will be shown how novel cavity design can achieve large couplings by ensuring the cavity field is primarily contained within the magnetic sample volume. It can also be seen this result that the fundamental limit of the coupling rate given a fixed cavity frequency is the material properties, including magnetic moment and spin density, of the sample as: $g_{cm}\propto \sqrt{\frac{\mu}{g\mu_B}n_s}$.\\

Spherical geometries of the ferrimagnetic samples are often used given the symmetries of the system makes the inclusion of demagnetising fields in modelling much simpler \cite{walker1, walker2, walker3}. For arbitrary non-spherical geometries, the magnon mode shapes are typically not known. As such, coupling rates can not be calculated from first principals as above. It is expected, and found in previous work, that the coupling rates can be related in general to the magnetic filling factor, $\zeta_m$, of cavity field contained in the sample by $g_{cm}^2=\omega_c^2\chi_{eff}\zeta_m$, where $\chi_{eff}$ is an effective susceptibility determined by material properties and the overlap of the specific cavity and magnon modes \cite{CQED1}. As the focus of this work is on non-spherical ferrite geometries, magnetic filling factor becomes the relevant figure of merit in maximising coupling rates. With the lack of predicted magnon frequency dependence with DC magnetic field in arbitrary geometries a phenomenological approach is taken. Typically a linear fit to magnon frequency is sufficient for at least part of the fitting procedure, where it can be related to the case of a spherical geometry governed by the Zeeman effect\cite{blundell}:

\begin{equation}
\omega_m(B_{DC}) = (2\pi)\frac{g_\text{eff}\mu_B}{\hbar}(B_{DC}+B_\text{off}),
\label{lineFit}
\end{equation}

where $\mu_B$ is the Bohr magneton, $B_\text{off}$ is an offset field typically due to magneto-crystalline anisotropies and $g_\text{eff}$ is the effective Land\'e g-factor, typically $g_\text{eff}=2$ for spherical geometries.

\section{Ultra strong coupling between magnetostatic mode of a YIG block and a re-entrant cavity}
\subsection{Cavity modelling and system specifications}
A rectangular prism of YIG, with dimensions 5$\times$3$\times$5 mm, containing a central 1 mm diameter hole, was placed inside a rectangular cavity. This cavity, with dimensions 5.5$\times$3.3$\times$5.2 mm, had a central re-entrant post of diameter 1 mm and height 5.1 mm. The cavity was placed inside a DC magnetic field oriented in the direction of the re-entrant post. Re-entrant cavities typically consist of a cylindrical cavity with a central post. The lowest order mode of this cavity has its electric field primarily between the post and the lid of the cavity; similar to a parallel plate capacitor, and the magnetic field around the post. Thus it forms a 3D lumped element LC resonator. Prior to measurement electromagnetic modelling is performed on the cavity, where the YIG block was considered a linear homogeneous dielectric with relative permittivity $\epsilon_r=15.96$\cite{USCmagnons}. This is shown in figure \ref{YIGcolor} (A) for several cavity modes with the primary mode of interest being labelled (1). As the YIG block takes up most of the cavity, it has a high magnetic filling factor of $\zeta_m=0.94$. Without knowledge of the specific magneto-static mode shape of the block, this is the primary design requirement to achieve large coupling between the two subsystems. Additionally as the gap between post and lid is free space, the electric filling factor in the block is low $\zeta_e=0.08$. This should reduce dielectric losses.

\subsection{Experimental Results and Discussions}

\begin{figure}[h!]
	\includegraphics[width=0.8\columnwidth]{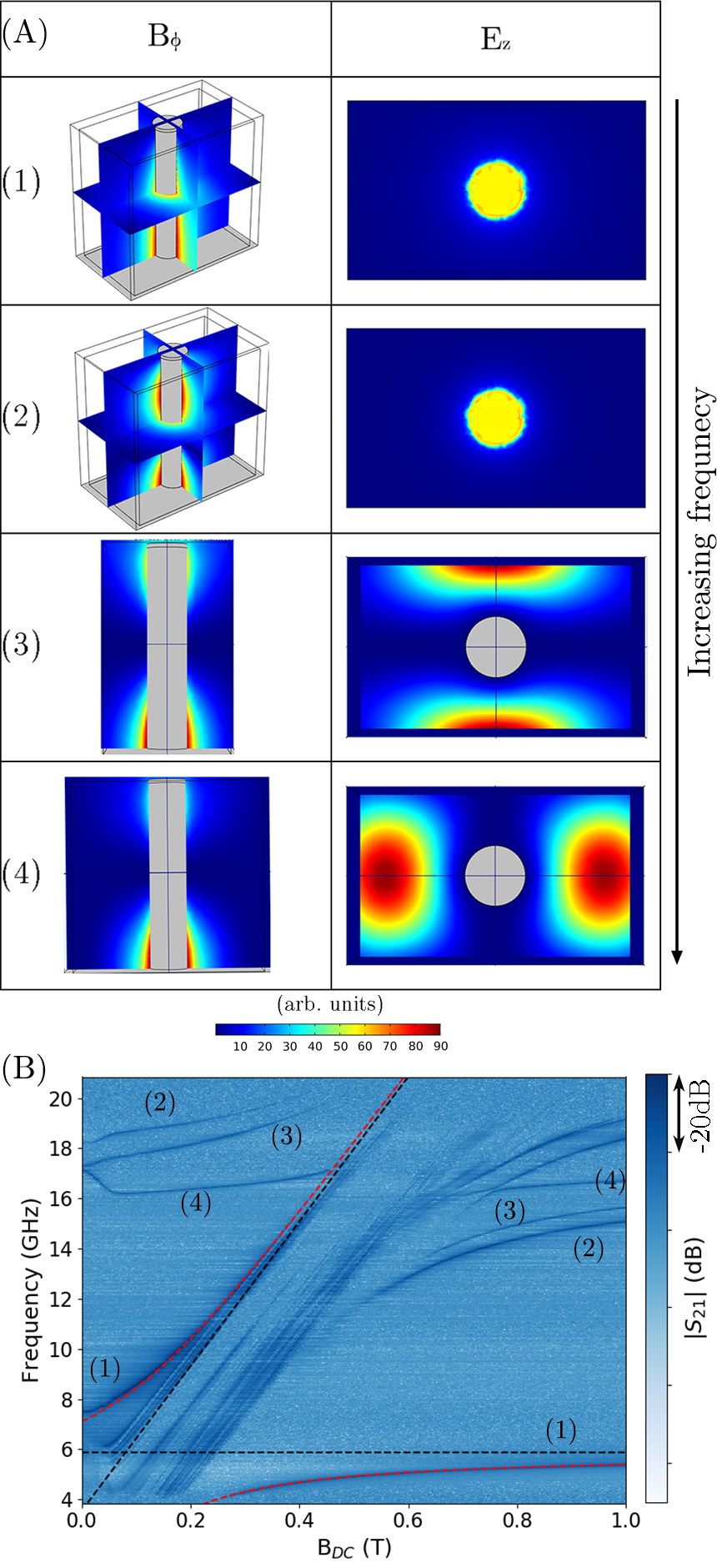}
	\caption{(A) Colour density plot of magnetic field strength (side view to the post) and electric field strength (plane transverse to the post) for the cavity modes of the re-entrant cavity. (B) YIG block system response in terms of its $S-21$ parameter as a function of external magnetic field with labels corresponding to the cavity modes. The red and black dashed lines are the fits to CMPs and uncoupled cavity and magnon modes associated with (1) respectively.}
	\label{YIGcolor}
\end{figure}

Spectroscopic data was taken with the system at approximately 20 mK, to probe the hybrid frequencies as a function of DC magnetic field and fitting was performed to the most strongly coupled hybrid modes by equation~(\ref{2mode}). The results of the spectroscopy and fitting are shown in figure~\ref{YIGcolor}. The fitted model parameters are $\omega_c/(2\pi) = 5.870\pm0.004$ GHz, $g_\text{eff} = 2.061\pm0.003$, $B_\text{off} = 0.1231 \pm 0.0003$ T, $g_{cm}/(2\pi) = 2.690 \pm 0.005$ GHz. The condition for ultra strong coupling is that the bare frequencies of the uncoupled modes satisfy the inequality: $g/\omega\geq0.1$. This is true of the cavity mode, as $g_{cm}/\omega_c = 0.46$, and the magnon mode, based on the assumed model, for $B_{DC}\leq0.81$ T.\\

Additional cavity modes can be seen in figure~\ref{YIGcolor} (B) exhibiting strong coupling to other magnon modes in the block. For completeness, the cavity modes predicted were identified by their frequencies (which should correspond to the measured frequencies at large field) and labelled. Their mode shape based on the electromagnetic modelling is shown in figure~\ref{YIGcolor} (A). It can be seen that modes (1) and (2) are the first and second order re-entrant post modes\cite{higherreentrant} as these have all the electric field in the gap between the post and lid. (3) and (4), on the other hand, appear to be perturbed rectangular cavity modes that would be degenerate if the cavity had x-y symmetry. Figure~\ref{YIGcolor} also shows that at low fields the modes, particularly the high frequency ones, have a sudden change in slope. This has been observed in the past in both YIG\cite{USCmagnons}, and yttrium aluminium garnet (YAG)\cite{YAG} and can be explained by the effect on the ferromagnetic phase of the impurity ions on degenerate modes.\\

It is unlikely that by further engineering the cavity, the coupling to this block couple be improved as it already had a near unity filling factor, however, with a material with higher spin density, such as LiFe, the coupling rate to this mode could potentially improve. Assuming the relation between coupling and spin density is $g_{cm}\propto \sqrt{\frac{\mu}{g\mu_B}n_s}$ as in the previous section, the coupling for an identical LiFe block in this cavity can be estimated. The ratio of spin density of LiFe relative to YIG is $\frac{n_{s,LiFe}}{n_{s,YIG}}\approx2.13$ where we note both YIG and LiFe have the same magnetic moment, $\mu$\cite{CMP_Life}. This is expected to improve couplings by a factor of $\sqrt{2.13}\approx1.46$. Thus the expected coupling with a LiFe block is $g_{cm}/(2\pi)\approx3.93$ GHz giving $g_{cm}/\omega_c \approx 0.67$.

\section{Ultra strong coupling between magnetostatic mode of a YIG disc and a loop gap cavity}
\subsection{Cavity modelling and system specifications}
A thin, single-domain YIG disc of diameter 6 mm and thickness 0.5 mm is placed in the central chamber of a loop gap cavity. Like the re-entrant cavity, the loop gap cavity is also a 3D lumped element LC resonator. In this case two rectangular cavities of dimensions 15$\times$35$\times$17 mm are separated by a copper slab of thickness 5 mm. The slab has a central cut-out region for the sample and two cylindrical cut-outs of 5 mm diameter with axes perpendicular to the ample axis, separated by 15.7 mm. The cylinders and central chamber form lumped element inductors with the magnetic field circulating around them. Two thin cut-out rectangular chambers between the cylinders and central chamber of dimensions 0.16$\times$2.35 mm, form lumped element capacitors with the majority of the electric field in these gaps. Thus they form two coupled 3D lumped element resonators. They hybridise to two normal modes, one where the electric fields of each oscillate in phase typically named the dark mode as it has little magnetic field in the central chamber, and the other where the electric fields oscillate out of phase forming the bright mode. The bright mode is characterised by the majority of the magnetic field directed through the central chamber with the sample and thus should achieve large magnetic filling factors. Once again, electromagnetic modelling is performed to determine the structure of the cavity modes. The bright mode has a large magnetic filling factor in the sample of $\zeta_m=0.77$. The electric filling factor in the disc is found to be low $\zeta_e=0.08$ ensuring low dielectric losses. 

\subsection{Experimental Results and Discussions}

\begin{figure}[h]
	\includegraphics[width=0.9\columnwidth]{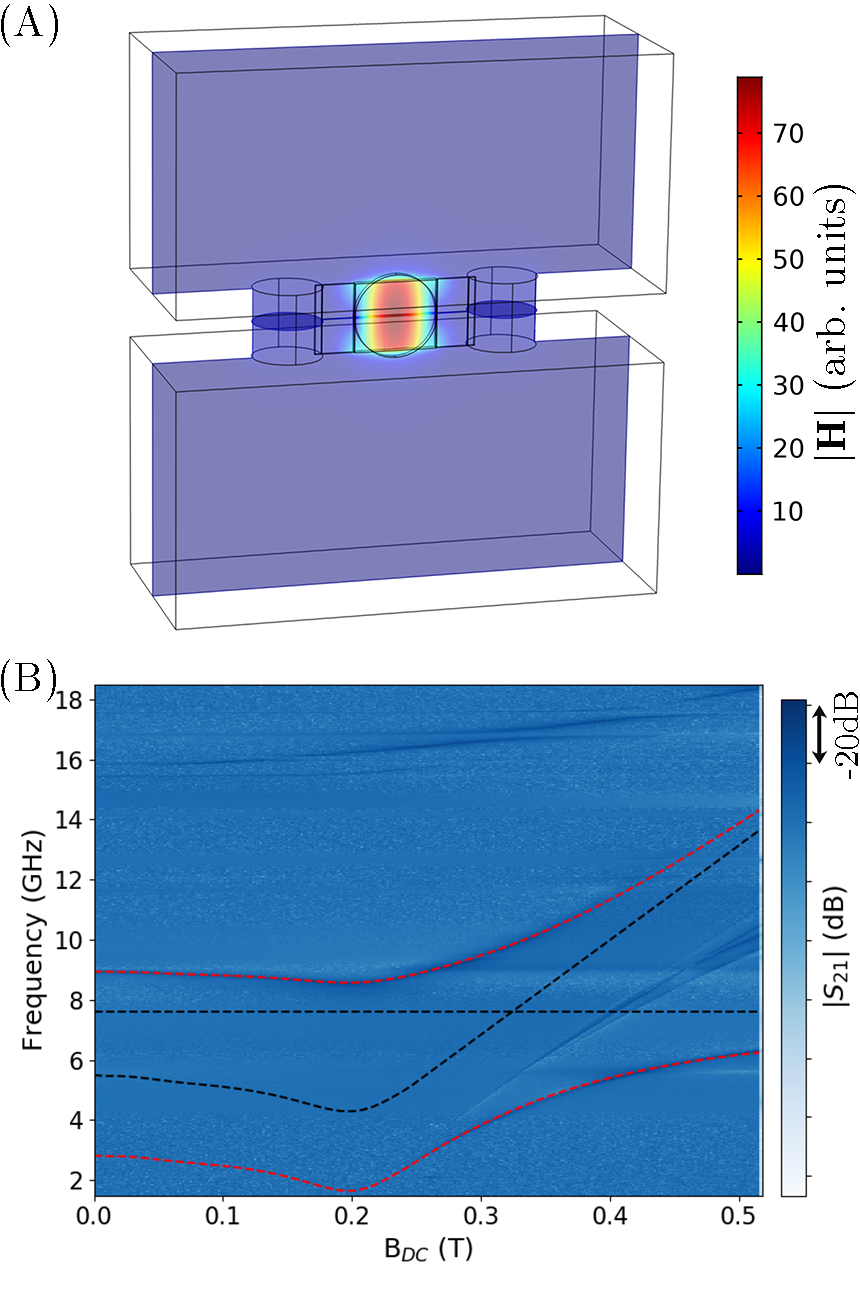}
	\caption{(A) Colour density plot of magnetic field strength $|\mathbf{H}|$ for the bright mode of the loop gap cavity. (B) YIG disc system response in terms of its $S-21$ parameter as a function of external magnetic field. The red and black dashed lines are the fits to CMPs and uncoupled cavity and magnon modes respectively.}
	\label{DISCcolor}
\end{figure}

Spectroscopic data was taken with the system at approximately 4 K to probe the hybrid frequencies as a function of DC magnetic field and fitting was performed to the most strongly coupled hybrid mode by equation~(\ref{2mode}). In this case a linear relationship with DC magnetic field was assumed where both branches of the hybrid system were visible ($B>0.27$ T approximately). An initial fitting was performed in this region to obtain the values of the cavity frequency and coupling rate. A polynomial relation to DC field was then used to infer the magnon frequency outside of this region assuming the coupling rate remains constant. The results of the spectroscopy and fitting are shown in figure~\ref{DISCcolor}. The fitted model parameters from the linear magnon fit are $\omega_c/(2\pi) = 7.599\pm0.008$ GHz, $g_\text{eff} =2.249\pm0.008$, $B_\text{off} = -0.083 \pm 0.001$ T, $g_{cm}/(2\pi) = 2.574 \pm 0.002$ GHz. The polynomial fit to the magnon frequency is shown in figure~\ref{DISCcolor} (B). The fitted parameters imply that the ultra-strong coupling regime is achieved as the ratio of coupling rate to cavity frequency is $g_{cm}/\omega_{c} = 0.34$ and the ratio to magnon frequency is inferred by the fitted parameters to be greater than 0.1 for $B\leq0.9$ T. \\

Again, as the filling factor was near unity, it is also unlikely that further cavity engineering will improve the measured coupling rate. The coupling can be improved with the use of the higher spin density medium of LiFe, as in the previous section. If an identical LiFe disc was used, the coupling can be expected to be $g_{cm}\approx(2\pi)3.76$ GHz giving $g_{cm}/\omega_c \approx 0.50$. If the measured magnon mode in question is assumed to be approximately a uniform precession mode, the expected coupling rate can be calculated by equation~(\ref{uniform}). This predicts a coupling of $g_{cm}/(2\pi)= 6.7$ GHz based on calculated form factors by the electromagnetic cavity model of $\eta=0.82$. This is obviously much larger than the coupling measured, thus, assuming the validity of equation~(\ref{uniform}), the limiting factor of this set-up is likely to be due to non-uniformity of the magnon mode leading to a suboptimal overlap with the cavity field. It would be potentially interesting, thus, to test a sphere inside a loop gap, where the mode shape is known to be uniform and coupling rates can be predicted. Assuming the cavity field is uniform and only in a cylindrical central chamber of a loop gap cavity with a sphere of arbitrary size in the centre, the form factor can be estimated as $\eta=0.82$. Thus with a cavity frequency of $\omega_c/(2\pi)= 5.9$ GHz, the expected coupling rate is $g_{cm}/(2\pi)\approx5.9$ GHz thus producing a ratio of $g_{cm}/\omega_c\approx1$, and reaching the deep strong coupling regime. \\

Cavity-magnon systems have been used in the past as methods for direct detection of axion dark matter in the form of ferromagnetic axion haloscopes\cite{me,quaxnew,quaxprop}. The interaction of the grad of the expected axion field with electron spins is analogous to that of a uniform oscillating magnetic field at the Compton frequency of the axion field \cite{origmagnonaxion, pseudoB1, pseudoB2}. It can be noted that the magnetic field in the loop gap cavity is approximately uniform and only in one direction. Thus the strong interaction of this field with the magnon mode suggests that the magnon mode would also make a prime candidate for dark matter detection. Without specific knowledge of the magnon mode structure, a first principles model of the axion-magnon interaction can't be determined. However, the system's response to uniform oscillating magnetic fields could be calibrated with a field of known size, thus inferring its interaction with axions. In the past, ferromagnetic axion haloscopes, have focused on spherical geometries for their magnetic material, as the magnon mode structure is well known. Non-spherical geometries can be beneficial from a practical standpoint as samples with larger volumes are often easier to acquire, where the number of spins in the system is a key parameter in increasing experimental sensitivity. Large cavity-magnon couplings lead to a larger range of frequencies and hence axion masses that the experiment can be sensitive to\cite{me}. Thus the USC achieved here would be advantageous for such an experiment to explore a larger range of the, as yet, unknown axion mass parameter space\cite{Irastorza:2018aa}.

\section{Applications of Ferrites in Frequency Metrology}
The focus of frequency metrology is typically in the development of high accuracy and stability clocks \cite{freqstand1, freqstand2, freqstand3}. However, applications of frequency metrology are also in fundamental physics\cite{metfunphys1, metfunphys2, metfunphys3, metfunphys4, axionexcept, axionMetrology}, including detection of dark matter through exceptional points \cite{axionexcept}. Exceptional points are achieved by engineering loss rates and couplings in open systems and increase a system's sensitivity to small perturbations in frequency. These have recently been demonstrated in cavity-magnon polaritons by adjusting the position of a small YIG sphere in a rectangular cavity \cite{magnonexcept}, as well as being investigated in the context of magnon-induced transparency and amplification \cite{PTsymmetricMagnon}. Additionally, in the context of dark matter detection, it was recently suggested that frequency metrology techniques applied to cavity-magnon polaritons could be used to detect ultralight axion or axion-like dark matter, as this interaction can appear as a modulation of magnon frequency\cite{me}.\\

In the commercial sector, an application of frequency metrology is in synthesisers and oscillators. In this context, for microwave frequencies, ferrites are particularly useful. High stability oscillators need large quality factors, something which microwave cavities in the form of sapphire whispering gallery mode resonators excel at \cite{highQ}. These devices are typically not compact, however. Ferrimagnetic thin films have significant advantage here, as they have modestly large quality factors and low phase noise, whilst potentially being able to be miniaturised \cite{filtersOscillators}. For broad applications, it is also useful to be widely tunable in frequency. As the frequency of ferromagnetic resonance (FMR) is determined primarily by an external DC magnetic field through the Zeeman effect, this allows a broad range of frequency tunability in such devices. As such, even spherical geometries find significant use as oscillators \cite{YIGmini1, YIGsphereosc1, YIGsphereosc2}. The high sensitivity of FMR frequency to DC magnetic field, whilst extremely useful for frequency tuning, also makes it extremely sensitive to magnetic field fluctuations, limiting the applicability of these devices in developing high stability devices. Thus, developing systems which are insensitive to these fluctuations would be of interest.

\section{LiFe sphere for reduced sensitivity to bias magnetic field fluctuations}

\subsection{Cavity-Magnon Polariton Transition Frequency fluctuations}
For any mapping of one variable to another, small fluctuations in the input variable will map to small fluctuations in the output by power series expansion. In our case we are interested in the conversion of fluctuations in the bias magnetic field, $\delta B$, to fluctuations in the CMP transition frequency, $\omega_\text{CMP}$, as follows, to second order:
\begin{equation}
\delta\omega_\text{CMP} = \deriv[B]{(\omega_\text{CMP})} \delta B + \derivtwo[B]{(\omega_\text{CMP})} \delta B^2 + O[\delta B^3].
\end{equation}
As such to suppress this conversion we are interested in suppressing the first and second derivative of CMP transition frequency response. A RWA can be used to simplify equation~(\ref{2mode}), giving the CMP transition frequency:

\begin{equation}
\omega_\text{CMP} = 2 \sqrt{\Big(\frac{\omega_{c}-\omega_{m}}{2}\Big)^2 + g_{cm}^2},
\end{equation}
where $\omega_{m}\equiv\omega_{m}(B)$. Thus the first and second derivatives are:
\begin{equation}
\deriv[B]{\omega_\text{CMP}} = -\frac{2(\omega_{c}-\omega_{m})}{\omega_\text{CMP}}\deriv[B]{\omega_{m}},
\label{firstderiv}
\end{equation}

\begin{multline}
\begin{aligned}
\derivtwo[B]{\omega_\text{CMP}} = &-\frac{2(\omega_{c}-\omega_{m})}{\omega_\text{CMP}}\derivtwo[B]{\omega_{m}} - \frac{2}{\omega_\text{CMP}}\Big(\deriv[B]{\omega_{m}}\Big)^2\\
&-\frac{4(\omega_{c}-\omega_{m})^2}{\omega_\text{CMP}^3}\Big(\deriv[B]{\omega_{m}}\Big)^2.
\end{aligned}
\label{secondderiv}
\end{multline}
In strongly coupled systems a turning point will naturally exist in the difference frequency of normal modes (ie. $\deriv[B]{\omega_\text{CMP}}=0$), when the difference in uncoupled mode frequencies is at a local minimum (or maximum) corresponding to a local maximum (minimum) in energy exchanged by the underlying degrees of freedom. This occurs when, either, the system is fully hybridised ($\omega_c=\omega_m$) or when the uncoupled modes also have a turning point ($\deriv[B]{\omega_{m}}=0$, where $\deriv[B]{\omega_c}=0$ is always true) as seen in equation (\ref{firstderiv}). When both of these conditions are met in the same system configuration, they correspond to an inflection point as seen in equation (\ref{secondderiv}) giving the desired suppression of bias field fluctuations to second order (ie. $\derivtwo[B]{\omega_\text{CMP}}=0$ and $\deriv[B]{\omega_\text{CMP}}=0$). A turning point in the magnon frequency's magnetic field dependence is, therefore, required.\\

Simple understanding of linear kittel magnon modes rely on the symmetry of single domain spheres, isotropy of spins in the sample and bias field above saturation, however, this is generally not the case in reality. Anisotropic and demagnetising fields are produced by magneto-crystalline anisotropy and breaking spherical symmetry of the sample respectively. These effects create a preferred direction of magnetization: the easy axis. When the applied magnetic field is below the saturation field, and is misaligned with this easy axis, the magnetization vector tends towards aligning with the easy axis rather than the applied field. Additionally, below the saturation field, multi-domain structures can develop. This mode softening behaviour combined with the normal Zeeman effect is the physical reason for the observed turnover point in magnon frequency. This is described in more detail for the sample measured\cite{CMP_Life} and in general\cite{turninguUniaxial, turningTheory} in the literature. Magneto-crystalline anisotropy has also been demonstrated to produce a magnon-Kerr non-linearity in YIG \cite{KerrMagnon, bistabilityMagnons, sidebandGenMagnons, MagnonChaos}. Given this anisotropy has also been measured in LiFe \cite{LiFeAnisotropy}, it is expected to also have a similar non-linearity. However, given the effect is expected to be extremely small, requiring large numbers of excitations to be visible, it can be neglected for this work.

\begin{figure}[h]
	\includegraphics[width=0.95\columnwidth]{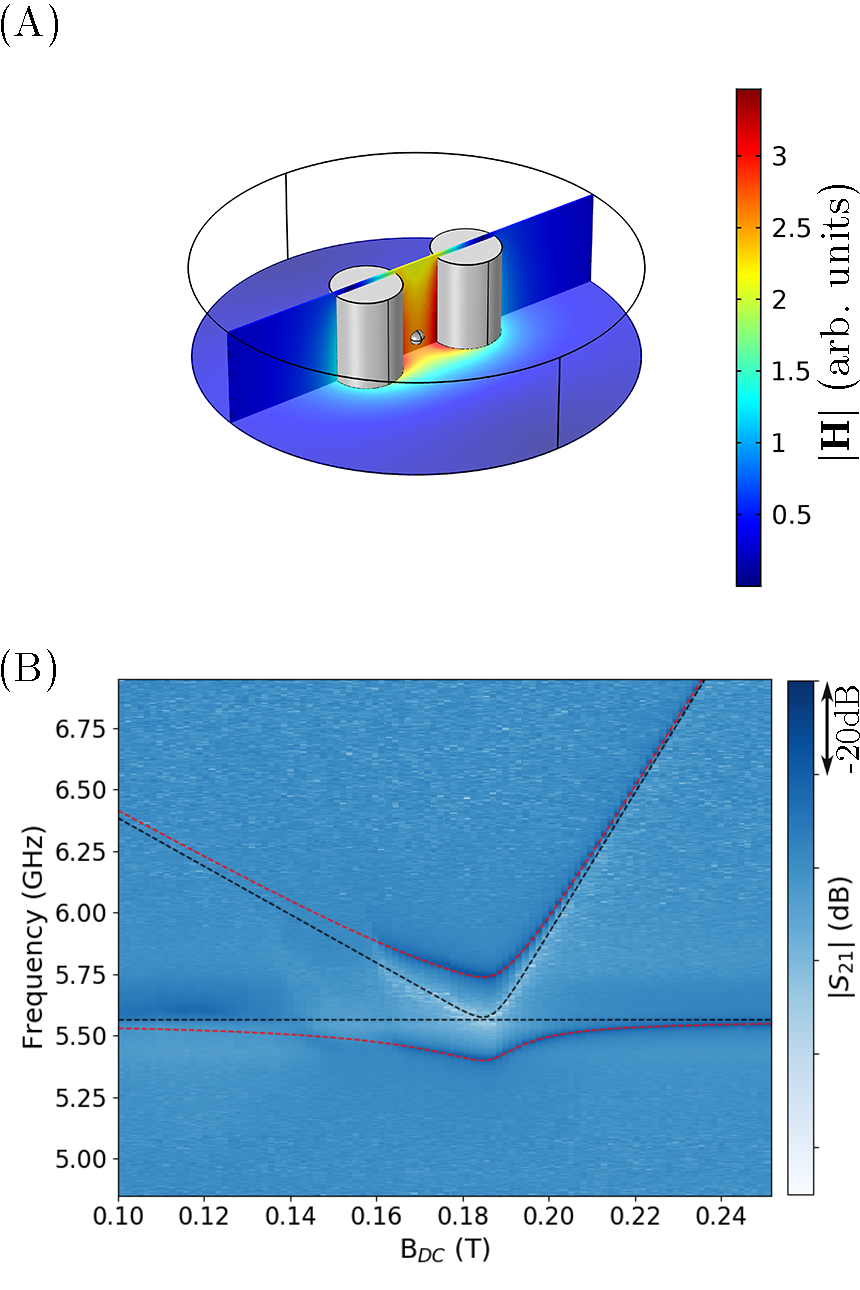}
	\caption{(A) Colour density plot of magnetic field strength $|\mathbf{H}|$ for the bright mode of the two post re-entrant cavity. (B) LiFe system response in terms of its $S-21$ parameter as a function of external magnetic field. The red and black dashed lines are the fits to CMPs and uncoupled cavity and magnon modes respectively.}
	\label{LiFe_color}
\end{figure}
\begin{figure}[h]
	\includegraphics[width=0.90\columnwidth]{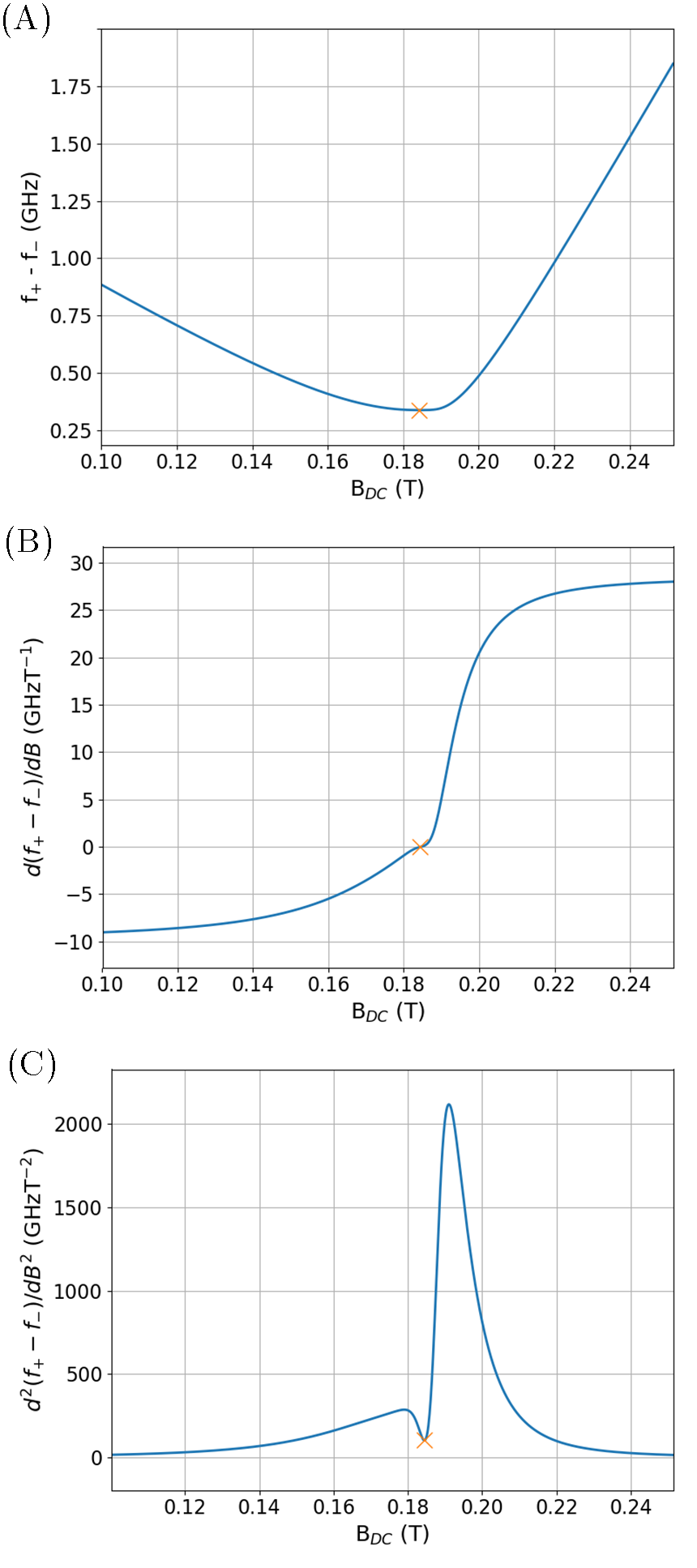}	
	\caption{CMP transition frequency as a function of DC magnetic field, (A), and its first, (B), and second, (C), derivatives. The orange cross marks the fully hybridized regime.}
	\label{LiFe_derivs}
\end{figure}
\subsection{Cavity modelling and system specifications}
The single domain, polished, LiFe sphere had a diameter of 0.58 mm. It is placed in a two post re-entrant cavity. This cavity is a cylinder of diameter 17.5 mm and height 3.8 mm. The two posts are of diameter of 2.8 mm, height of 3 mm and separation of 4.8 mm from centre to centre. A small spacer of 0.1 mm was inserted between the cavity walls and lid. This allowed cavity frequency tuning through fine adjustments to the gap size between the posts and cavity lid by tightening the lid screws. This once again forms two coupled 3D lumped element resonators. The applied DC magnetic field is oriented along the direction of the post and the (110) crystal axis of the LiFe sample. The mode of interest is the bright mode where the majority of the magnetic field are directed between the posts. This is described in more detail in previous works \cite{CMP_Life,CQED1}. The electromagnetic modelling for this mode is shown in figure \ref{LiFe_color} (A). 

\subsection{Experimental Results and Discussions}
The fit given by equation (\ref{lineFit}) is applied to the linear sections of the spectroscopy data with a phenomenological fit applied to the turnover point. The results of the spectroscopy with the system at approximately 20 mK, including applied fits, is shown in figure~\ref{LiFe_color}. This gives fitting parameters: $g_\text{eff,p} = 2.03$, $B_\text{off,p} = 0.00780$ T, $g_\text{eff,m} = -0.70$, $B_\text{off,m} = -0.751$ T, $\omega_c/(2\pi) = 5.56$ GHz, $g_{cm}/(2\pi) = 169$MHz, where the subscripts p and m refer to the positive and negative sloped limits of the hybrid mode frequency respectively.\\

The fitting in figure~\ref{LiFe_color} (B), shows the cavity frequency is close to the turnover point in frequency as designed. The parameter of interest is the CMP transition frequency (difference frequency of hybrid modes). This parameter and its first and second derivatives are shown in figure~\ref{LiFe_derivs} where the point of maximal hybridisation, corresponding to the minimum in CMP transition frequency, is marked with a cross. It can be seen that the CMP transition is insensitive to first order bias magnetic field fluctuations when maximally hybridised as usual, however it can also be seen that the second order field fluctuations are also suppressed.  Continued tuning of the cavity frequency such that it is closer to this turnover point will reduce this sensitivity further. If an oscillator were constructed based on this transition frequency, its frequency stability would be reduced by fluctuations of the bias field. This is a significant limitation of these devices. Thus by appropriately operating the demonstrated system as an oscillator at the maximally hybridised bias point, the frequency stability is expected to be improved.

\section*{Conclusions}
In conclusion, we have presented three implementations of cavity-magnon experiments. The first two focussed on the implementation of strong coupling between cavity and magnon degrees of freedom through cavity engineering. Non-spherical ferrite geometries were specifically investigated and phenomenological fits to spectroscopic data matched well with the measured data. Ultra strong coupling was shown to be achieved in both systems making them prime candidates for use in the study of cavity QED or hybrid quantum systems. It is expected that the limiting factor in maximising couplings in the presented systems was the spin density of the material and non-ideal mode overlaps. The use of a loop gap cavity was found to be particularly interesting for future investigation, as the magnetic field in the bright mode of this cavity is also consistent with the use of spherical ferrimagnetic samples where mode overlaps are expected to be better. It was predicted that the deep strong coupling regime should be reasonably achievable via these cavities. Additionally, it is expected to make a good candidate for improved axion dark matter detection. The last system presented attempted to fully hybridise a two post re-entrant cavity mode with a magnon mode of a LiFe sphere at its turnover point in bias magnetic field. This was successful and demonstrated a suppression in the sensitivity of the CMP transition frequency to both first order and second order fluctuations in bias magnetic field, making it useful for applications in frequency metrology as an oscillator.

\section*{Acknowledgements}
This work was supported by the Australian Research Council grant number DP190100071 and CE170100009 as well as the Australian Government's Research Training Program.

\section*{References}
\bibliography{refs}

\begin{figure*}[t!]
	\begin{minipage}{2.1\columnwidth}\flushleft
		\textsf{\textbf{\Large Experimental Implementations of Cavity-Magnon Systems: from Ultra Strong Coupling to Applications in Precision Measurement - Supplementary Material}}
	\end{minipage}
\end{figure*}
\newpage

\section{Cavity-Magnon coupling assuming a uniform precession mode}
The Hamiltonian of the cavity-magnon system consists of its cavity and magnon parts, $H_c$ and $H_m$ respectively, as well magnon-cavity interaction, $H_\textrm{int}$:
\begin{multline}
\begin{aligned}
H &= H_c + H_m + H_\textrm{int}\\
H &= \hbar\omega_c c^\dagger c + \hbar\omega_m b^\dagger b + H_{int},
\end{aligned}
\end{multline}
where $c^\dagger$ ($c$) is a creation (annihilation) operator for photon, $b^\dagger$ ($b$) is a magnon creation (annihilation) operator, $\omega_c$ is the cavity frequency, $\omega_m$ is magnon frequency and $\hbar$ is the reduced Planck's constant. These expressions can been found from first principles \cite{cavitymagnon, magnonkerr}. The interaction term can then be evaluated by the Zeeman energy of the ferrimagnetic sample:
\begin{align}
H_{int} &= -\mu_0\int_{V_m}\mathbf{M}\cdot\mathbf{H}_{c}\mathrm{d}V,
\label{zeeman}
\end{align}
where $\mathbf{M}$ is the magnetisation, $\mathbf{H}$ is the cavity mode auxiliary magnetic field, $\mu_0$ is the permeability of free space and the integral is performed over the magnetic sample volume $V_m$. The form of the magnetic field in the cavity mode can be found by the usual methods of quantisation:
\begin{align}
\mathbf{H}_{c} &= \frac{1}{\mu_0}\sqrt{\frac{\hbar}{2\omega_c\epsilon_0\epsilon_{r,c}}} (c +c^\dagger)\mathbf{\nabla}\times\mathbf{U},
\label{Hquantum}
\end{align}
where $\epsilon_0$ is the permittivity of free space and $\epsilon_{c,r}$ is an average relative permittivity experienced by the cavity mode defined as:
\begin{align}
\epsilon_{r,c} &= \int_{Vc}\epsilon_r\mathbf{U}\cdot\mathbf{U}\mathrm{d}V. 
\end{align}
Finally, $\mathbf{U}$ solves the wave equation and is orthonormal with other cavity modes as follows, respectively:
\begin{align}
\frac{1}{\epsilon_r}\nabla\times\nabla\times \mathbf{U}_{n} - \frac{\omega_n}{c^2}\mathbf{U}_{n} &=0, \\
\int_{Vc}\mathbf{U}_{n}\cdot\mathbf{U}_{m}\mathrm{d}V &= \delta_{nm},
\end{align}
where $c$ is the speed of light and $\delta_{nm}$ is the Kronecker delta.\\

The uniform precession mode of the magnetic sample can also be quantised by introducing a macrospin $\mathbf{S}=\frac{\mathbf{M}V_m}{\gamma}$. If we assume the direction of the DC magnetic field which saturates the magnetic material is in the z direction we can then introduce raising and lowering operators ($S^{\pm}=S_x\pm iS_y$) followed by the Holstein-Primakoff transformations \cite{primakoffHolstien}:
\begin{multline}
\begin{aligned}
S^+ &= (\sqrt{2S - b^\dagger b})b,\\
S^- &= b^\dagger(\sqrt{2S - b^\dagger b}),\\
S_z &= S - b^\dagger b,
\end{aligned}
\end{multline}
where $S$ is the total spin number of the macrospin operator. This number is determined by $S=\frac{\mu}{g\mu_B}N_s$, where $\mu$ is the magnetic moment of the magnetic sample, $\mu_B$ is the Bohr magneton, $g$ is the g-factor ($g=2$) and $N_s$ is the number of spins in the sample (given by $N_s=n_sV_m$ with $n_s$ as spin density and $V_m$ as volume). For YIG, for example, it is estimated that $\frac{\mu}{\mu_B}=5.0$\cite{blundell} and the spin density $n_s=2.2\times10^{28}$m$^{-3}$.\\

For low excitation numbers (${\langle b^\dagger b\rangle}\ll 2S$), the macrospin operators may be approximated by $S^+\approx\sqrt{2S}b$ and $S^-\approx\sqrt{2S}b^\dagger$. If we substitute eqn.~\ref{Hquantum} and these transformations into eqn.~\ref{zeeman}, we arrive at the interaction Hamiltonian:

\begin{multline}
\begin{aligned}
H_{int}/\hbar = &g_{cm}^{x}(c+c^\dagger)(b+b^\dagger) + g_{cm}^{y}(c+c^\dagger)(b-b^\dagger)\\
&+ g_{cm}^{z}(c+c^\dagger)b^\dagger b + \Omega^z(c+c^\dagger)
\label{Hint}
\end{aligned}
\end{multline}
where the coupling rates are defined as:
\begin{multline}
\begin{aligned}
g_{cm}^{x} &= -\frac{\gamma}{2V_m}\sqrt{\frac{\hbar S}{\omega_c\epsilon_{r,c}\epsilon_0}}\int_{V_m}(\nabla\times\mathbf{U})\cdot\hat{x}\mathrm{d}V,\\
g_{cm}^{y} &= \frac{i\gamma}{2V_m}\sqrt{\frac{\hbar S}{\omega_c\epsilon_{r,c}\epsilon_0}}\int_{V_m}(\nabla\times\mathbf{U})\cdot\hat{y}\mathrm{d}V,\\
g_{cm}^{z} &= \frac{\gamma}{V_m}\sqrt{\frac{\hbar}{2\omega_c\epsilon_{r,c}\epsilon_0}}\int_{V_m}(\nabla\times\mathbf{U})\cdot\hat{z}\mathrm{d}V,\\
\Omega^{z} &= -\frac{\gamma S}{V_m}\sqrt{\frac{\hbar}{2\omega_c\epsilon_{r,c}\epsilon_0}}\int_{V_m}(\nabla\times\mathbf{U})\cdot\hat{z}\mathrm{d}V.\\
\end{aligned}
\end{multline}
To aid in the evaluation of these expressions the following relations can be used:
\begin{multline}
\begin{aligned}
\mathbf{U} &= \frac{\mathbf{E}}{\sqrt{\int_{V_c}|\mathbf{E}|^2\mathrm{d}V}},\\
\nabla\times\mathbf{U} &= \frac{\sqrt{\epsilon_{r,c}}\omega_c}{c} \frac{\mathbf{H}}{\sqrt{\int_{V_c}|\mathbf{H}|^2\mathrm{d}V}}.
\end{aligned}
\end{multline}
These expressions, which are normalised to cavity energy, allow the use of calculated field shapes to evaluate the coupling rates. The first two terms in the interaction hamiltonian are standard coupled mode terms due to cavity RF field perpendicular to the external DC field coupling to the magnon mode. This will be the focus of this analysis. The third term is a parametric term due to the cavity RF field parallel to the external DC field modulating the magnon frequency and the final term is a result of the DC saturation magnetisation generating field in the cavity. These last two terms can be neglected under a rotating wave approximation or by ensuring the cavity field is perpendicular to the external DC field. That is the case in this work. \\

\end{document}